\documentclass{article}

\usepackage []{graphicx,color}
\usepackage{authblk}
\usepackage{natbib}
\usepackage{amsmath}

\title{Doubly stochastic  distributions of extreme events}

\author[1,2]{Marco Marani \thanks{corresponding author, marco.marani@unipd.it}}
\author[2]{Enrico Zorzetto}
\affil[1]{Universit{\`a} degli studi di Padova, Dipartimento di Ingegneria Edile, Civile ed Ambientale (ICEA)}
\affil[2]{Duke University, Division of Earth and Ocean Sciences}

\begin{document}
\maketitle

\begin{abstract}
The distribution of block maxima of sequences of independent and identically-distributed random variables is used to model extreme values in many disciplines. The traditional extreme value (EV) theory derives a closed-form expression for the distribution of block maxima under asymptotic assumptions, and is generally fitted using annual maxima or excesses over a high threshold, thereby discarding a large fraction of the available observations. The recently-introduced Metastatistical Extreme Value Distribution (MEVD), a non-asymptotic formulation based on doubly stochastic distributions, has been shown to offer several advantages compared to the traditional EV theory. In particular, MEVD explicitly accounts for the variability of the process generating the extreme values, and uses all the available information to perform high-quantile inferences. Here we review the derivation of the MEVD, analyzing its assumptions in detail, and show that its general formulation includes other doubly stochastic approaches to extreme value analysis that have been recently proposed.
\end{abstract}

\section{Introduction}
Classical extreme value (EV) theory starts from considering a process of event occurrences in time \citep{fisher1928limiting, von1936distribution, gnedenko1943distribution, coles1996modelling, de2007extreme}. Each event occurrence is characterized by a magnitude (referred to as {\it ordinary event} magnitude in the following). Ordinary event magnitudes are assumed to be realizations from independent and identically distributed (i.i.d.) random variables. Next, the time axis is divided into blocks of equal duration. In the following we focus, without loss of generality, on the case of yearly blocks, which is common in many applications in hydrology and environmental sciences. EV theory then derives the probability distribution of the maximum value, $m$, among the magnitudes $\left\{ x_1, \cdots,  x_n \right\}$ of all the $n$ events occurring in the same year. The yearly maximum is the realization of a new random variable $M$, whose distribution is sought. Note the variable values, $n$, of the number of ordinary events in a given block (e.g. the case of non-zero daily rainfall events), which can also be viewed as the realizations of a discrete random variable, $N$.

    The traditional EV theory derives the distribution of the annual maxima either as the asymptotic distribution in the limit for $n \rightarrow \infty$, or through the assumption of Poisson-distributed exceedances of a high threshold. The result of the traditional EV theory is the Generalized Extreme Value (GEV) distribution, which summarizes the possible forms of extreme value distributions arising under the above assumptions   \citep{fisher1928limiting, von1936distribution, gnedenko1943distribution,  balkema1974residual, pickands1975statistical, leadbetter1983extremes, smith1984threshold, davison1990models}. The issue of convergence to the limting EV distribution has been widely investigated, and is known to be particularly slow for certain parent distribution commonly used in EV applications \citep{cook2004exact, harris2006errors, lugrin2019penultimate}.
    Additionally, while classical EV theory provides a closed-form expression for the cumulative distribution function of $M$, its application discards most of the available observations. Furthermore, the GEV distribution does not explicitly address those cases in which the underlying distribution generating the ordinary values may vary across years, either due to systematic trends or to interannual climatic fluctuations.
    
\citet{marani2015metastatistical} and \citet{zorzetto2016emergence} recently introduced and extensively tested a non-asymptotic distribution for the yearly maxima of i.i.d. sequences of ordinary events, termed the \emph{Metastatistical Extreme Value Distribution} (MEVD). The MEVD is a doubly stochastic approach \citep{dubey1968compound, beck2003superstatistics} that explicitly accounts for i) the random nature of the number of events/year and ii) the inter-annual variability of the distributions of the ordinary events in each year. The approach was first applied to daily rainfall amounts, showing that the MEVD significantly improves over traditional approaches in estimating extreme rainfall values when the number of years of observations ($S$) is small compared to the return time ($T_r$) of the extreme event being estimated
\citep{zorzetto2016emergence}. Further investigations extended the approach to hourly rainfall \citep{marra2018metastatistical} and to rainfall remote sensing estimates \citep{zorzettoandmarani2019}, showing, in particular, the robustness of the method with respect to observational uncertainty and short sample sizes. 

A recent contribution \citep{de2018superstatistical}  introduced a superstatistical extreme value distribution, also a doubly stochastic approach to extreme value analysis.
In the following, we analyze similarities and differences of these two formulations, which we find to be originated from the restrictive assumption in \citet{de2018superstatistical} that the yearly number of events be Binomial-distributed. In the process, we also clarify the nature, relevance, and limitations of doubly stochastic extreme value models.

\section{Occurrence process and dependence structure of event magnitudes}

The basic assumption of classical extreme value theory is that the random variables representing the magnitudes associated with each event are i.i.d., with a common cumulative probability function $P(X < x \mid \vec{\theta}) = F(x \mid \vec{\theta})$ ($\vec{\theta}$ being the vector of the parameters defining the distribution). As a result of this assumption, the cumulative distribution of the yearly maximum, $P ( M \leq x \mid \vec{\theta}, n)$, can be written as the n-th power of the ordinary-event cumulative distribution, expressing the probability that all $n$ magnitudes in a year are smaller or equal to $x$:
\begin{equation}
P \left( M \leq x \mid \vec{\theta}, n \right) = P \left( X \leq x \mid \vec{\theta} \right)^n = F(x \mid \vec{\theta})^n
\label{MEVD_annual}
\end{equation}
In the case of intermittent phenomena, such as rainfall accumulations, the \emph{events} are usually defined as the non-zero accumulations $x \in (0, +\infty)$, so that their number in a given year is not fixed and can therefore be described by a random variable $N$. The distribution of the annual maximum is then obtained by noting that, for each of the $n$ events, there is a probability $F(x \mid \vec{\theta})$ that the magnitude be smaller or equal to $x$, and a probability $1-F(x \mid \vec{\theta})$ that the magnitude be greater than $x$. Hence, the probability distribution of the k-th order statistic, $x_1 < x_2 < ... < x_k < ... < x_n$, or the probability that $x_k \leq x$ can be written as \citep{kottegoda2008applied}:
\begin{equation}
 P (x_k < x \mid \vec{\theta}, n) = F_k(x \mid \vec{\theta})= \sum_{j=k}^{n}  F(x \mid \vec{\theta})^k \left[1- F(x \mid \vec{\theta})\right]^{n-k}
\label{binomialF}
\end{equation}
When $k=n$ one obtains the cumulative distribution of the yearly maximum $P (x_n < x \mid  \vec{\theta}, n)=P(M \leq x \mid \theta, n)= F_n(x \mid \theta )=F(x \mid \vec{\theta})^n$. Note that this result does not require any hypothesis on the event arrival process generating the value $n$, and that the probability considered in eq. (\ref{binomialF}), $F(x \mid \vec{\theta})$, is conditional to the occurrence of an event, i.e. is defined for $x>0$. In other words, eq. (\ref{binomialF}) is valid for any probability distribution of $n$.

%A different model could be constructed by defining as ordinary events all values in $\Omega_H = [0, +\infty)$, with distribution $H(x, \mid \theta)$. In the case of an intermittent process, such as daily rainfall, $H(x, \mid \theta)$ contains an atom of probability in zero. In this case, the probability distribution of the annual maximum is $P(M<x|N_t,\vec{\theta}) = H(x \mid \theta)^{N_t}$, where $N_t = 365$ is the constant number of trials. This expression can be obtained only under the assumption that all daily occurrences are independent of each other, and as a particular consequence the arrival of non-zero events must follow a binomial distribution.%

% In other words, even though eq. (\ref{binomialF}) may mistakenly seem to be based on a binomial distribution of the number of events in a year, no assumption is made other than the {\bf magnitudes be i.i.d.}. 

%In the following we carefully examine the relation between these models, together with a numerical example and the application to observed time series.

\section{The Metastatistical Extreme Value Distribution}
\citet{marani2015metastatistical} propose to consider $\vec{\theta}$ as random parameter vector, to capture inter-annual and/or systematic climatic variability in the ordinary rainfall distribution, and define the MEVD as the ensemble average of the probability distributions of the maximum in each year over the variability of $\vec{\theta}$ and $n$:
\begin{equation}
    \zeta_{MEVD} (x) =  P \left( M \leq x \right) = \sum_{n = 0}^{\infty} \int_{\Omega_{\vec{\theta}}} F(x \mid \vec{\theta})^n g(n,\vec{\theta}) d\vec{\theta}
    \label{MEVD}
\end{equation}
where $g(n,\vec{\theta})$ is the joint probability distribution of the parameter vector and of the number of wet days. $\Omega_{\vec{\theta}}$ denotes the population of $\vec{\theta}$. \citet{marani2015metastatistical} also propose to evaluate the MEVD as an average on the available $S$-year samples of $N$ and $\vec{\theta}$, thereby avoiding the need to impose an analytical expression for $g(n,\vec{\theta})$:
\begin{equation}
    \zeta_{MEVD}(x) \simeq \frac{1}{S}\sum_{i = 1}^{S}{ \left[F(x, \vec{\theta_i})\right]^{n_i}
}
\end{equation}
\citet{marani2015metastatistical} and \citet{zorzetto2016emergence}, who focus on daily rainfall, assume the distribution of ordinary events, $F(x \mid \theta)$, to be a two-parameter Weibull distribution, such that $\vec{\theta} = \left\{ C, w\right\}$, where $C$ and $w$ denote Weibull's scale and shape parameters. The Weibull CDF reads:
\begin{equation}
    F(x \mid C, w) = 1 - e^{ - (x/C) ^w}
\end{equation}
A similar assumption is made by \citet{marra2018metastatistical} for hourly rainfall (after proper decorrelation of the hourly values) and by \citet{zorzettoandmarani2019} in the case of daily rainfall remote sensing estimates. In general, the specific choice of the probability distribution of the magnitude of ordinary events will depend on the characters of the process at hand. In the following we use a general notation that can be specialized for different applications through different choices of $F(x)$.

\section{\citet{de2018superstatistical}}

\citet{de2018superstatistical} represent the intermittent daily rainfall process as a bivariate sequence of random variables
$ \left\{ \left( J_n , X_n \right) , n \geq 0 \right\}$. The first random variable $J_n$, describing the process of event arrival, is initially assumed to be a two-state Markov Chain 
characterized by transition probabilities $p_{ij} = P \left[ J_n = j| J_{n-1} = i \right] $ with $\left\{ i,j\right\} = \left\{ 0,1\right\}$ 
characterizing dry or wet conditions respectively. The random variables $X_n$ are conditionally independent, given the state of $J_n$, as in the traditional extreme value theory. 
Their cumulative distribution functions are $F_i(x) = P\left( X \leq x | J_n = i\right)$, where $F_1(x)$ is assumed to be Weibull, as in \citet{marani2015metastatistical}, whereas $F_0(x)$ is degenerate, i.e. it is a finite atom of unit probability at zero.

\citet{de2018superstatistical} do not find a closed-form distribution of annual maxima  from the above first-order Markov chain assumption and introduce a further restrictive hypothesis, i.e. that the occurrence process be a zero-order Markov chain. Note that this is equivalent to assuming {\bf independence in the event occurrence process} in addition to the hypothesis of independence among the ordinary non-zero magnitudes made in the derivation of the MEVD and of the traditional EV theory. This additional assumption in \citet{de2018superstatistical} implies a binomial distribution of event occurrence, an assumption that is not needed for eq. (\ref{binomialF}) and hence for MEVD (or the traditional extreme value theory). When a binomial process of event occurrence is assumed, the distribution of the ordinary event magnitudes, including the zeroes, is:
\begin{equation}
F(x \mid p_0, \vec{\theta}) = p_0 + (1-p_0)F_1(x, \vec{\theta})
\label{DA18_1}
\end{equation}
 where $p_0$ is the "probability of zero rainfall in a day" \citep{de2018superstatistical} and $F_1(x, \vec{\theta})$ is the distribution of non-zero rainfall. Using this expression in eq. (\ref{MEVD_annual}), where now the number of 'events' is fixed and equal to the number of days in a year, leads to the following expression for the distribution of the annual maximum
\begin{equation}
    F_M(x \mid p_0, \vec{\theta}) = \left[ p_0 + (1-p_0)F_1(x, \vec{\theta})\right]^{N_t}
    \label{DA18_annual}
\end{equation}
where $N_t = 365$ days.
 Subsequently, similarly to \citet{marani2015metastatistical}, \citet{de2018superstatistical}) assume both the
 distributional parameters, $\vec{\theta}$, and $p_0$ to vary from year to year in a metastatistical fashion. This leads to the following expression for the average cumulative distribution of annual maxima:
\begin{equation}
    \zeta_{DA18}(x) = \frac{1}{S}\sum_{i = 1}^{S}{ \left[ p_{0_i} + (1-p_{0_i})F_1(x, \vec{\theta_i})\right]^{N_t}
        \label{annual_sevd}
}
\end{equation}
Where, again following \citet{marani2015metastatistical}, the distribution of the magnitudes of 'ordinary' rainfall events, $F_1(x \mid \vec{\theta})$, is taken to be Weibull with parameters $\vec{\theta} = \left\{ C, w\right\}$. As a point of departure from the MEVD approach, \citet{de2018superstatistical} introduce an additional threshold parameter, as a Weibull position parameter $\mu$, which is determined by minimizing the distance (as measured by a Kolmogorov-Smirnov statistic) between the distribution (\ref{annual_sevd}) and the cumulative frequency of the observed annual maxima.

%In the following section we examine differences and similarities between this approach and the aforementioned MEVD. 

\section{The MEVD as the general distribution of yearly maxima}

In further support of the notion that eq. (\ref{annual_sevd}) is a MEVD in which a binomial event occurrence process is assumed, we recover here eq. (\ref{DA18_annual}) from the MEVD formulation.

We start from the definition of the MEVD, eq. (\ref{MEVD}), where initially, with \citet{de2018superstatistical}, distributional parameters are assumed to be fixed:
\begin{equation}
\zeta (x) = \sum_{n = 0}^{\infty}p(n) F(x \mid C, w)^n
\label{MEVDsimple}
\end{equation}

 Next we assume, as in \citet{de2018superstatistical}, that the distribution of the number of wet days in a year is binomial with parameter $p_0$:
\begin{equation}
p(n) = \binom{N_T}{n} p_0^{N_T - n} \left( 1-p_0 \right)^{n}
\label{binomial}
\end{equation}
 By substituting this expression into eq. (\ref{MEVDsimple}), one obtains:
\begin{equation}
\zeta (x) = \sum_{n = 0}^{\infty}  \quad F(x \mid C, w)^n \binom{N_T}{n} p_0^{N_T - n} \left( 1-p_0 \right)^{n} 
\end{equation}

Which can be rearranged to give:
\begin{equation}
\zeta (x) =  \sum_{n = 0}^{\infty} \binom{N_T}{n} \quad p_0^{N_T - n} \left[  F(x \mid C, w) \left( 1-p_0 \right) \right]^n 
\label{almostthere}
\end{equation}
by recalling that:
\begin{equation}
\sum_{n = 0}^{\infty} \binom{N_T}{n} x^{N_T - n} y^n = (x+y
)^{N_T} 
\end{equation}
eq. (\ref{almostthere}) becomes:
\begin{equation}
\zeta (x) = \left[ p_0 + \left(1-p_0 \right) F(x \mid C, w)\right]^{N_T}
\end{equation}
Which is eq. (\ref{DA18_annual}) and the basis for the derivation of the so-called superstatistical distribution of extremes described by \citet{de2018superstatistical}. Hence, we conclude that the formulation for the distribution of annual maxima derived in \citet{de2018superstatistical} can be obtained from \citet{marani2015metastatistical} and \citet{zorzetto2016emergence} once the restrictive hypothesis is made that $p(n)$ be a binomial distribution. \citet{de2018superstatistical} test the hypothesis that event occurrences be serially correlated. We note that, upon analyzing data from 21,510 stations from the GHCN dataset, they find that "a first-order Markov chain seems more appropriate to represent the observed time series than the simpler case of zero-order Markov chain". The latter assumption is, however, chosen "for simplicity".

\section{Conclusions}
The analyses discussed here show that the MEVD is a general doubly stochastic and non-asymptotic extreme value distribution. In particular, the results in \citet{de2018superstatistical} can be retrieved as a special case of the MEVD. The main assumption in the derivation of the MEVD (as in the traditional extreme event theory) is that the magnitudes of ordinary events be iid, whereas no particular temporal structure of wet event arrivals is assumed.  % While the work by DA18 presents some novel elements, such as the introduction of a tunable threshold, these should have been presented and evaluated with respect to the existing theory. 

\clearpage
\bibliographystyle{agufull08}
\bibliography{rainfallbib} 

\begin{thebibliography}{21}
\providecommand{\natexlab}[1]{#1}
\expandafter\ifx\csname urlstyle\endcsname\relax
  \providecommand{\doi}[1]{doi:\discretionary{}{}{}#1}\else
  \providecommand{\doi}{doi:\discretionary{}{}{}\begingroup
  \urlstyle{rm}\Url}\fi

\bibitem[{\textit{Balkema and De~Haan}(1974)}]{balkema1974residual}
Balkema, A.~A., and L.~De~Haan (1974), Residual life time at great age,
  \textit{The Annals of probability}, pp. 792--804.

\bibitem[{\textit{Beck and Cohen}(2003)}]{beck2003superstatistics}
Beck, C., and E.~G. Cohen (2003), Superstatistics, \textit{Physica A:
  Statistical mechanics and its applications}, \textit{322}, 267--275.

\bibitem[{\textit{Coles and Tawn}(1996)}]{coles1996modelling}
Coles, S.~G., and J.~A. Tawn (1996), Modelling extremes of the areal rainfall
  process, \textit{Journal of the Royal Statistical Society. Series B
  (Methodological)}, pp. 329--347.

\bibitem[{\textit{Cook and Harris}(2004)}]{cook2004exact}
Cook, N.~J., and R.~I. Harris (2004), Exact and general ft1 penultimate
  distributions of extreme wind speeds drawn from tail-equivalent weibull
  parents, \textit{Structural Safety}, \textit{26}(4), 391--420.

\bibitem[{\textit{Davison and Smith}(1990)}]{davison1990models}
Davison, A.~C., and R.~L. Smith (1990), Models for exceedances over high
  thresholds, \textit{Journal of the Royal Statistical Society. Series B
  (Methodological)}, pp. 393--442.

\bibitem[{\textit{De~Haan and Ferreira}(2007)}]{de2007extreme}
De~Haan, L., and A.~Ferreira (2007), \textit{Extreme value theory: an
  introduction}, Springer Science \& Business Media.

\bibitem[{\textit{De~Michele and Avanzi}(2018)}]{de2018superstatistical}
De~Michele, C., and F.~Avanzi (2018), Superstatistical distribution of daily
  precipitation extremes: A worldwide assessment, \textit{Scientific reports},
  \textit{8}(1), 14,204.

\bibitem[{\textit{Dubey}(1968)}]{dubey1968compound}
Dubey, S.~D. (1968), A compound weibull distribution, \textit{Naval Research
  Logistics Quarterly}, \textit{15}(2), 179--188.

\bibitem[{\textit{Fisher and Tippett}(1928)}]{fisher1928limiting}
Fisher, R.~A., and L.~H.~C. Tippett (1928), Limiting forms of the frequency
  distribution of the largest or smallest member of a sample, in
  \textit{Mathematical Proceedings of the Cambridge Philosophical Society},
  vol.~24, pp. 180--190, Cambridge Univ Press.

\bibitem[{\textit{Gnedenko}(1943)}]{gnedenko1943distribution}
Gnedenko, B. (1943), Sur la distribution limite du terme maximum d'une serie
  aleatoire, \textit{Annals of mathematics}, pp. 423--453.

\bibitem[{\textit{Harris}(2006)}]{harris2006errors}
Harris, R. (2006), Errors in gev analysis of wind epoch maxima from weibull
  parents, \textit{Wind and Structures}, \textit{9}(3), 179--191.

\bibitem[{\textit{Kottegoda and Rosso}(2008)}]{kottegoda2008applied}
Kottegoda, N.~T., and R.~Rosso (2008), \textit{Applied statistics for civil and
  environmental engineers}, Blackwell Malden, MA.

\bibitem[{\textit{Leadbetter}(1983)}]{leadbetter1983extremes}
Leadbetter, M.~R. (1983), Extremes and local dependence in stationary
  sequences, \textit{Probability Theory and Related Fields}, \textit{65}(2),
  291--306.

\bibitem[{\textit{Lugrin et~al.}(2019)\textit{Lugrin, Davison, and
  Tawn}}]{lugrin2019penultimate}
Lugrin, T., A.~C. Davison, and J.~A. Tawn (2019), Penultimate analysis of the
  conditional multivariate extremes tail model, \textit{arXiv preprint
  arXiv:1902.06972}.

\bibitem[{\textit{Marani and Ignaccolo}(2015)}]{marani2015metastatistical}
Marani, M., and M.~Ignaccolo (2015), A metastatistical approach to rainfall
  extremes, \textit{Advances in Water Resources}, \textit{79}, 121--126.

\bibitem[{\textit{Marra et~al.}(2018)\textit{Marra, Nikolopoulos, Anagnostou,
  and Morin}}]{marra2018metastatistical}
Marra, F., E.~I. Nikolopoulos, E.~N. Anagnostou, and E.~Morin (2018),
  Metastatistical extreme value analysis of hourly rainfall from short records:
  Estimation of high quantiles and impact of measurement errors,
  \textit{Advances in Water Resources}, \textit{117}, 27--39.

\bibitem[{\textit{Pickands~III et~al.}(1975)}]{pickands1975statistical}
Pickands~III, J., et~al. (1975), Statistical inference using extreme order
  statistics, \textit{the Annals of Statistics}, \textit{3}(1), 119--131.

\bibitem[{\textit{Smith}(1984)}]{smith1984threshold}
Smith, R.~L. (1984), Threshold methods for sample extremes, in
  \textit{Statistical extremes and applications}, pp. 621--638, Springer.

\bibitem[{\textit{Von~Mises}(1936)}]{von1936distribution}
Von~Mises, R. (1936), La distribution de la plus grande de n valeurs,
  \textit{Rev. math. Union interbalcanique}, \textit{1}(1).

\bibitem[{\textit{Zorzetto and Marani}(2019)}]{zorzettoandmarani2019}
Zorzetto, E., and M.~Marani (2019), Downscaling of rainfall extremes from
  satellite observations, \textit{Water Resources Research}, \textit{55},
  \doi{10.1029/2018WR022950}.

\bibitem[{\textit{Zorzetto et~al.}(2016)\textit{Zorzetto, Botter, and
  Marani}}]{zorzetto2016emergence}
Zorzetto, E., G.~Botter, and M.~Marani (2016), On the emergence of rainfall
  extremes from ordinary events, \textit{Geophysical Research Letters},
  \textit{43}(15), 8076--8082.

\end{thebibliography}
\end{document}